\documentstyle[aps,prl,epsfj,multicol]{revtex}
\draft

\newcommand{\be}{\begin{equation}}
\newcommand{\ee}{\end{equation}}
\newcommand{\bd}{\begin{displaymath}}
\newcommand{\ed}{\end{displaymath}}
\newcommand{\ba}{\begin{eqnarray}}
\newcommand{\ea}{\end{eqnarray}}
\newcounter{fgr}
\def\figline#1#2{\vbox{
\refstepcounter{fgr}
\label{fig.\thefgr}
\vskip 0.5cm
\centerline{\mbox{\epsfile{file={#1},width=9.cm,height=5.5cm}}}
\vskip -3pt \hskip 0.cm
\centerline{\mbox{\hskip -0.5cm
\vtop{\hsize=8cm \noindent\strut{\bf Fig.\thefgr}\strut \hskip 0.3cm{\small
#2}} }}
}
\par}

\def\av#1{{\left\langle#1\right\rangle}}
\def\pht#1#2{\partial#1/\partial#2}
\def\rf#1{[\ref{#1}]}

\def\Rl{R_{\lambda}}\def\etal{{\em et al.}}\def\ub{{\bar{u}}}
\def\eb{{\bar{\epsilon}}}
\def\r{{\mbox{\boldmath$r$}}}\def\x{{\mbox{\boldmath$x$}}}
\def\Pic{{\mit{\Pi}}}

\begin{document}

\title{\vbox to 0pt {\vskip -1cm \rlap{\hbox to \textwidth {\rm{\small}
\hfill 
}}}Pressure spectrum and structure function in homogeneous 
turbulence}

\author{{\bf Toshiyuki Gotoh$^{1}$ and Daigen Fukayama$^{2}$}}

\address{$^1$ Department of Systems Engineering, Nagoya Institute of
Technology, \\ Showa-ku, Nagoya, 466-8555, Japan}
\address{$^2$ Department of Physics, Chuo University, Tokyo, 112-8551, Japan}
\date{Version 1.0, \today}

\maketitle

\begin{abstract}\noindent
The pressure spectrum and structure function in 
homogeneous steady turbulence of an incompressible 
fluid is studied using direct numerical simulation. 
The resolution of the simulation is up to $1024^3$ and 
the Taylor microscale Reynolds number $\Rl$ is between 
$38$ and $478$. The energy spectrum is found to have a small but finite 
inertial range followed by a bump at large wavenumbers. 
The Kolmogorov constant $K$ is found to be $1.66\pm 0.08$. 
The pressure spectrum also has a small but 
finite inertial range of $P(k)=B_p\eb^{4/3}k^{-7/3}$ followed by 
a bump of nearly $k^{-5/3}$ range at higher wavenumbers. 
Both scaling ranges match at a crossover wavenumber, $k_p$, which is 
a characteristic wavenumber for the pressure gradient. 
The constant $B_p$ is found to be 
about $8.34\pm 0.15$ for $\Rl=460$. 
Its non-universality is discussed. 
The second order pressure structure function, computed in terms of 
the fourth order velocity structure functions, agrees well with that
obtained by 
the direct measurement over separations ranging between 
the inertial and dissipation scales.
 
\end{abstract}

%\noindent Corresponding author: Toshiyuki Gotoh

%\noindent Department of Systems Engineering,

%\noindent Nagoya Institute of Technology,

%\noindent Showa-ku, Nagoya 466-8555, Japan

%\noindent Tel: +81-52-735-5377 \quad Fax: +81-52-735-5401

%\noindent Email: gotoh@system.nitech.ac.jp

\begin{multicols}{2}
The pressure spectrum in a turbulent incompressible flow is 
defined as $\av{p^2}=\int^{\infty}_0 P(k) dk$. Kolmogorov's theory 
predicts that 
\ba
    P(k)&=&\eb^{3/4}\nu^{7/4}\phi(k\eta)    \nonumber\\
        &=&B_p\eb^{4/3}k^{-7/3}, \quad L^{-1}\ll k\ll \eta^{-1}, \label{1}
\ea
where $\phi(x)$ is a non-dimensional function, $\nu$ is
the kinematic viscosity, 
$\eb$ is the average rate of energy dissipation per unit mass, $L$ is the 
integral scale of turbulence, $\eta$ is the Kolmogorov scale, and
$B_p$ is a non-dimensional constant of order one. 
The fluid density is assumed to be unity throughout this paper. 

There have been many studies of the pressure spectrum.
\cite{Bat51,MY75,Georgeetal84,Holzer93,Gr94,Pr94,Pumir94,Nelkin94,P95,Hw95,Hill96,Oaza96,Hb97,Hillthoro97,Vothetal98,Ved99,Caoetal99,Gr99,Gn99,Pa00} 
Some of the experiments have shown that 
$P(k)\propto k^{-7/3}$,\cite{Georgeetal84} or equivalently 
$D_p=\av{(p(\x+\r)-p(\x))^2}\propto r^{4/3}.$\cite{Oaza96,Pa00} 
Others have reported that $r^{4/3}$ is not observed.\cite{Hb97}
Recent DNS's with large scale forcing have found that 
the pressure spectrum is approximately proportional 
to $k^{-5/3}$, unlike $k^{-7/3}$, in the wavenumber range 
where the energy spectrum scales close to $k^{-5/3}$.\cite{Ved99,Caoetal99,Gr99}
Gotoh and Rogallo conjectured that 
the observed $k^{-5/3}$ scaling for $P(k)$ is a bump, as observed for
the energy spectrum, and that $P(k)$ scales as $k^{-7/3}$ 
in the lower wavenumber 
range.  This implies that a wider inertial range, compared to the energy 
spectrum, 
is necessary for K41 scaling 
of $P(k)$.\cite{Gr99}
There seems to be no agreement about the scaling of the pressure spectrum 
when compared to the case of the energy spectrum. 

We have performed a series of DNS's of incompressible 
homogeneous isotropic turbulence using a resolution of up to $N=1024^3$. 
The DNS was designed to generate a wider inertial range and 
higher Reynolds numbers. 
The range of the Taylor microscale Reynolds number $\Rl=\bar{u}\lambda/\nu$ 
is between $38$ and $478$, where $\bar{u}$ is the root mean square of 
turbulent velocity and $\lambda$ is the Taylor microscale. 
The characteristic parameters of the DNS are listed in Table I. 
The code uses the pseudo Fourier spectrum and 4th order Runge Kutta 
Gill methods. Random forcing, Gaussian and white in time, is applied to the 
lower wavenumbers. A statistically steady state was confirmed by 
observing the time evolution of the total energy, the total enstrophy and 
the skewness of the longitudinal velocity derivative. 
The statistical averages were 
taken as the time average over tens of turnover times for lower Reynolds 
numbers and over a few turnover times for the higher Reynolds numbers. 
The data of the highest Reynolds number, $\Rl=478$, were  
obtained as short time average (about 0.34 eddy turnover times)  
during the passage toward steady state rather than over
a statistically steady state. 
The resolution condition $k_{max}\eta>1$
is satisfied for most runs, but that of the case when $\Rl=460$ is slightly 
less than unity. We believe that this does not adversely affect the
energy and pressure spectra results in the inertial range. 
Computations with $\Rl\le 284$ were performed using a Fujitsu VPP700E
vector parallel 
machine with $16$ processors at RIKEN.  Simulations using 
higher $\Rl$ were performed on
a Fujitsu VPP5000/56 with $32$ processors 
at the Nagoya University Computation Center. 

The energy spectra in Kolmogorov units (multiplied by 
$(k\eta)^{5/3}$) are shown in Fig.\ref{fig.1}. 
Collapse of curves at various Reynolds numbers is very satisfactory, 
although the curves with $\Rl\ge 284$ have appreciable rise of $E(k)$ near 
the high wavenumber boundary. 
The Kolmogorov constant $K$ was measured in the range of 
$0.008\le k\eta\le 0.04$ in which the average energy transfer flux 
function $\Pic(k)/\eb$ is nearly flat and close to unity (figure not shown).
The value of $K$ 
\be
  K=1.66\pm 0.08,                                    \label{0}
\ee 
is very close to the value of $1.62$ obtained in previous experiments 
and DNS's,\cite{Sreeni95,Yz97}
and to the value of $1.72$ obtained using
the Lagrangian spectral theory (LRA).\cite{K81,K86} 
There is a small spectral bump at wavenumbers 
near $k\eta\approx 0.2$, as observed in other DNS's.\cite{Yz97}

Figure 2 shows the pressure spectra in terms of the K41 scaling, Eq.\rf{1},  
multiplied by $(k\eta)^{7/3}$ for various Reynolds numbers.
Fig. \ref{fig.3} is a close-up of the curves for higher Reynolds numbers. 
For $\Rl<300$ there is no plateau in the curves. However, 
for $\Rl$ larger than $400$, there appears to be a 
small plateau of finite length. 
There is a bump, with a peak value of about $17$, in
$P(k)$ near $k\eta=0.2$ which is more appreciable than 
that in the energy spectrum. 
The left part of the bump consists of a finite ramp.
For $\Rl=284$, the slope of the ramp 
is close to $2/3$, indicating that $P(k)\propto k^{-5/3}$.  The slope 
gradually decreases as the Reynolds number increases. 
It is this part which the previous 
DNS's have observed as $P(k)\propto k^{-5/3}$.\cite{Ved99,Caoetal99,Gr99} 

The curves obtained for $\Rl=387$, $460$ and $478$ indicate that $P(k)$ 
approaches the $k^{-7/3}$ spectrum over the range of $0.008<k\eta<0.04$. 
The value $B_p$ is 
\be
B_p=8.34\pm 0.15 ,                                     \label{01}
\ee
shown in Fig.3 as a horizontal line. 
Taking into account the relatively short length of the averaging time, 
the error bar for $B_p$ is a few times larger than $0.15$. 
Pullin obtained $B_p=1.325K^2$ using the joint Gaussian 
hypothesis for the 4th order velocity structure functions. 
With $K=1.66$, $B_p=3.65$, which is 
smaller than the present DNS value. 
This is consistent with the fact that $P(k)$ is larger than $P_G(k)$,
computed from the Gaussian random velocity field with the same energy 
spectrum as that of the actual turbulence field.\cite{Gr94,Pr94,Gr99} 
Pumir suggested $B_p\approx 7$ using DNS data with $N=128^3$ at $\Rl=77.5$.
Pullin estimated that $B_p\approx 2.14-3.65$ 
using Lundgren's stretched spiral 
vortex model. 

The collapse of the $P(k)$ curves for all $\Rl$'s is not as good as the 
energy spectrum, even in the dissipation range. 
The collapse of the pressure spectra is improved when
the normalized pressure gradient variance, 
$F_{\nabla p}=\av{(\nabla p)^2}\eb^{-3/2}\nu^{1/2}$, 
is included in the scaling for $P(k)$: 
\be
   P(k)=F_{\nabla p}\eb^{3/4}\nu^{7/4}\phi_1(k\eta),     \label{2}
\ee
where $\phi_1(x)$ is a non-dimensional function.\cite{Gn99,Gnw00}  
Figure \ref{fig.4} presents $P(k)$ using Eq.\rf{2}, 
and clearly shows that the scaling of $P(k)$ 
in the high wavenumber range is better than the scaling using Eq.\rf{1}. 
The inset shows the variation of $F_{\nabla p}$ against the Reynolds number. 
$F_{\nabla p}$ is a monotonically increasing 
function of $\Rl$ that becomes very weakly dependent on 
$\Rl$ as $\Rl$ becomes large. 
It should be noted that although the insensitivity of $F_{\nabla p}$ to $\Rl$ 
is consistent with Batchelor's Gaussian theory for the pressure,
its value is considerably larger than the value 
corresponding to the Gaussian theory, $F_{\nabla p}^G$.\cite{Bat51,Gr99} 
This insensitivity of $F_{\nabla p}$ at large $\Rl$ implies that 
the collapse of $P(k)$ with $\Rl\ge 284$ is little affected by 
$F_{\nabla p}$. However, there still remains a 
weak Reynolds number dependence of $P(k)$ in the inertial range, 
causing the pressure spectrum to be non-universal.
Close inspection of $P(k)$ in the $k^{-7/3}$ range 
shows that the factor $F_{\nabla p}$ of Eq.\rf{2} 
improves the collapse of the curves (Figs. \ref{fig.2} and \ref{fig.4}). 
In this range, $\phi_1(k\eta)=C_p(k\eta)^{-7/3}$, 
where $C_p$ is a non-dimensional 
constant of the order one. The constant $B_p$ is related as 
$B_p=F_{\nabla p}C_p$, so that $B_p$ becomes 
weakly dependent on Reynolds number, while $C_p$ is not. 
It is reasonable, in this sense, to regard $C_p$ as a more universal 
constant than $B_p$. Using the values of $B_p$ and $F_{\nabla p}$
we obtain 
\be
C_p=0.707\pm 0.1.              \label{2b}
\ee
%at around $\Rl=460$. 

The non-universality enters the pressure spectrum through $F_{\nabla p}(\Rl)$ 
as a function of $\Rl$. Therefore, $F_{\nabla p}$ is a key parameter 
for the second order statistics of pressure. 
The Reynolds number dependence of $F_{\nabla p}$ is attributed to 
the coherent structure of the source term field in 
the Poisson equation 
for the pressure.\cite{Gr99,Gnw00} 

Transition to the nearly $k^{-5/3}$ range occurs at 
$k\eta\approx 0.03$ for $\Rl\ge 387$, which corresponds to 
$k_p\lambda_p\approx 1.5, 1.8$ and $1.6$ for $\Rl=387, 460$ and $478$, 
respectively. Here, $\lambda_p$ is a characteristic length scale
for the pressure gradient defined by $\av{(\pht{p}{x})^2}=\ub^4/\lambda_p^2$. 
This is analogous to the Taylor microscale. 
Thus the crossover scale between the $k^{-7/3}$ and the 
nearly $k^{-5/3}$ ranges is about $k_p=\lambda_p^{-1}$. 

The pressure spectrum at low wavenumbers scales as 
\be
P(k)=\eb^{4/3}L^{7/3}\phi_2(kL),             \label{2c}
\ee
where $L$ is the integral scale. 
The curves of $P(k)$ at this range 
collapse reasonably well into one curve (figure not shown).  
The scaling of Eq.\rf{2} matches Eq.\rf{2c} in the $k^{-7/3}$ range.  

Hill and Wilczak have derived an expression for 
the pressure structure function 
$D_p(r)$ in terms of the 
fourth order velocity structure functions, assuming a homogenous and
isotropic velocity field:
\ba
D_p(r)&=&-{1\over 3}L(r)
          +{4\over 3}r^2\int^{\infty}_r\!\!y^{-3}[L(y)+T(y)-6M(y)]dy
                                                              \nonumber\\ 
      & & \hskip 0.5cm +{4\over 3}\int^{r}_0y^{-1}[T(y)-3M(y)]dy,  \label{6}
\ea
where $L(r)=\av{\delta u(r)^4}, T(r)=\av{\delta v(r)^4}$,
$M(r)=\av{\delta u(r)^2\delta v(r)^2}$, and $\delta u(r)$ 
and $\delta v(r)$ are the longitudinal and transversal velocity differences, 
respectively.\cite{Hw95}

Isotropy has been examined in terms of the kinematic 
conditions for the second and third order moments of 
the longitudinal and transversal velocity increments.
The relations are well satisfied for both $\Rl=387$ and $\Rl=460$. 
(The relative error in the equation 
for the third order moments is less than about $10\%$ at $r/\eta=200$.) 

Figure \ref{fig.5} shows plots of $L(r), T(r)$ and $M(r)$ at $\Rl=387$ and
$460$. 
There is a straight line contained along each curve between 
$50<r/\eta<400$. The straight line showes the slope of $1.28$, the value 
predicted by She and L\'ev\^eque.\cite{Sl94} 
An examination of the local exponent of the curves showed that
the inertial range exponents defined by 
$L(r)\propto r^{\zeta_4^L}, T(r)\propto r^{\zeta_4^T}$ and 
$M(r)\propto r^{\zeta_4^M}$ have plateau between $90<r/\eta<200$ and 
$1.24<\zeta^T_4<\zeta^M_4<\zeta^L_4<1.32$ for $\Rl=460$.\cite{Pa00}
The detailed analysis will be reported elsewhere.\cite{Kmg00} 

$D_p(r)$ computed by Eq.\rf{6} is compared with values obtained using direct 
measurement in Fig. \ref{fig.6}. The curves for $\Rl=460$ are shifted 
upward by one unit for clarity. Agreement of the curves for $r/\eta<150$ 
is satisfactory.
It is reasonable to expect that 
the isotropy of the fourth order velocity structure 
functions in the limit of $r\to 0$, the fundamental 
assumption for the derivation of \rf{6}, is well satisfied.
The pressure gradient variance and the pressure structure function 
at small separations can be examined in terms of Eq.\rf{6}.\cite{Hw95,Pa00} 
In the inset of Fig.\ref{fig.4}, the values of $F_{\nabla p}$ computed using 
$D_p(r)r^{-2}$ in the limit of $r\to 0$ are found to be very close 
to the value of
$F_{\nabla p}$ obtained by direct measurement. 

The straight line in Fig.\ref{fig.6} indicates the $r^{1}$ slope. 
The slope of $D_p(r)$ is very close to unity and between $2/3$ and $4/3$. 
The scaling of $D_p(r)$ requires a much 
longer scale separation than the case using wavenumber space. 

\vskip 0.2cm
The authors acknowledge Profs. Antonia and Nakano for their 
helpful discussions, and express their thanks to Nagoya University 
Computation Center and the Advanced Computing Center at RIKEN for providing
the computational resources. 
This work was supported by a Grant-in-Aid for Scientific Research 
(C-2 12640118) by The Ministry of Education, Science, Sports and Culture 
of Japan. T.G's work was supported by the "Research for 
the Future" Program of the Japan Society for the Promotion of Science,
under the project JSPS-RFTF97P01101.

\end{multicols}

\vspace*{-0.6cm} 

\begin{table*}
\caption{DNS parameters and Statistical quantities of runs: 
$T_{eddy}^{av}$ is the length of time average.}
\vspace*{0.2cm} 
%\begin{tabular}{cccccccccccccccc}
\begin{tabular}{cccccccc@{}dc@{$\!\!$}d@{\ \ \ }cddcc}
$\Rl$  &   $N$    & $k_{max}\eta$ &      $\nu$     & $c_f$ & forcing range
&$T^{av}_{eddy}$
& $E$ & $\epsilon$ & $L$ & $\lambda$ & $\lambda_p$ &\multicolumn{1}{c}{\ \
$\eta${\scriptsize ($\times 10^{-2}$)}}& $F_{\nabla p}$ & $K$ & $B_p$ \\
\hline
38  & 128$^3$ &    60     & $1.50\times10^{-2}$ &  1.30  & $\sqrt{3}\leq k \leq \sqrt{12}$
& 22.6 
&1.99 &1.19  &0.891 &0.501  & 0.371 & 4.10 & 3.62 & - & - \\
70 & 256$^3$ &    121    & $4.00\times10^{-3}$ &  0.50  & $\sqrt{3}\leq k \leq \sqrt{12}$
& 49.7
&1.16 &0.457 &0.785 &0.318  &0.256 & 1.93 & 5.60 & - & - \\
125  & 512$^3$ &    241    & $1.35\times10^{-3}$ &  0.50  & $\sqrt{3}\leq k \leq
\sqrt{12}$ & 5.52
&1.25 &0.492 &0.744 &0.185  &0.170 & 0.841 & 7.61 & - & - \\
284  & 512$^3$ &    241    & $6.00\times10^{-4}$ &  0.50  & $1\leq k \leq
\sqrt{6}$ & 3.03
&1.96 &0.530 &1.246 &0.149  &0.177 & 0.449 & 10.4 & 1.64 & - \\
387  & 1024$^3$ &   483    & $2.80\times10^{-4}$ &  0.51  & $1\leq k \leq
\sqrt{6}$  & 1.09 
&1.81 &0.522 &1.215 &0.0986 &0.131 & 0.255  & 11.3 & 1.62 & - \\
460  & 1024$^3$ &   483    & $2.00\times10^{-4}$ &  0.51  & $1\leq k \leq
\sqrt{6}$ & 1.43 
&1.79 &0.506 &1.150 &0.0841 &0.119 & 0.199  & 11.8 & 1.64 & 8.48 \\ 
478  & 1024$^3$ &   483    & $2.80\times10^{-4}$ &  0.51  & $1\leq k \leq
\sqrt{6}$ & 0.34
&2.00 &0.419 &1.350 &0.116  &0.142 & 0.269  & 11.8 & 1.74 & 8.19 \\ %\hline
\end{tabular}
\label{table1}
\end{table*}

\begin{multicols}{2}
\noindent

\vspace*{-0.4cm}

%%%%%% Spectrum %%%%%%
\figline{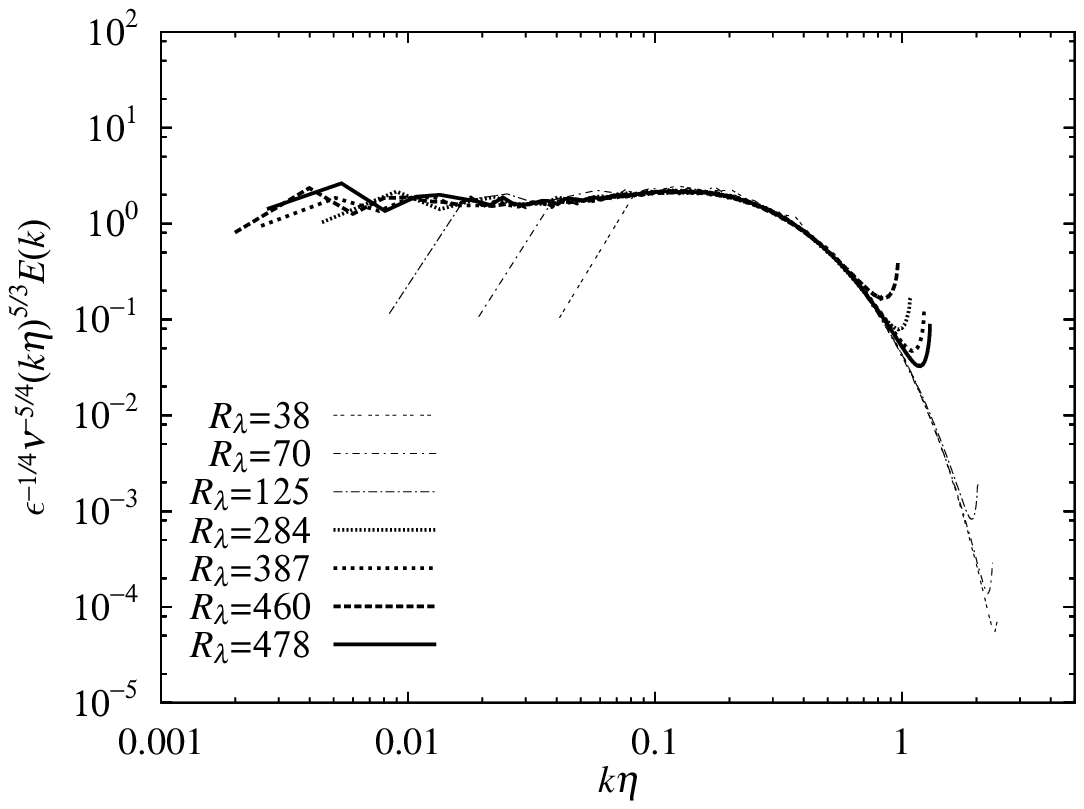}{Scaled energy spectra,
$\eb^{-1/4}\nu^{-5/4}(k\eta)^{5/3}E(k)$.  $K=1.66\pm 0.08$. }
\figline{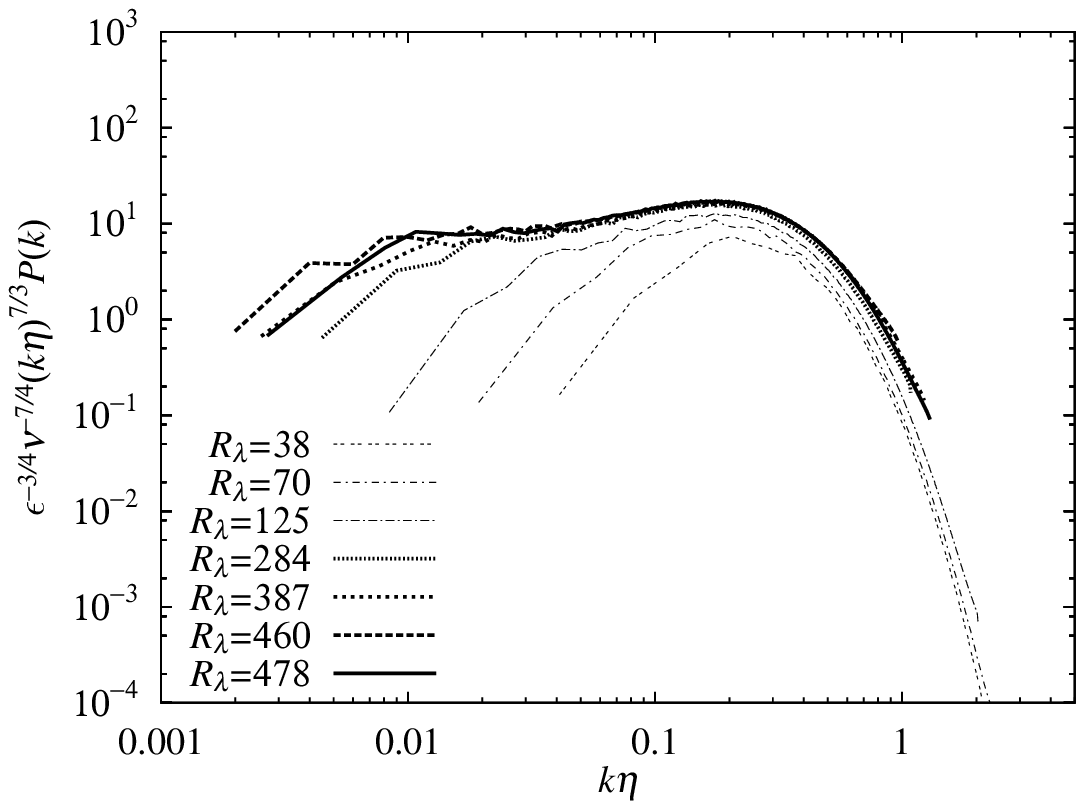}{K41 scaling for the pressure spectra,
$\eb^{-3/4}\nu^{-7/4}$ $(k\eta)^{7/3}P(k)$. }
\figline{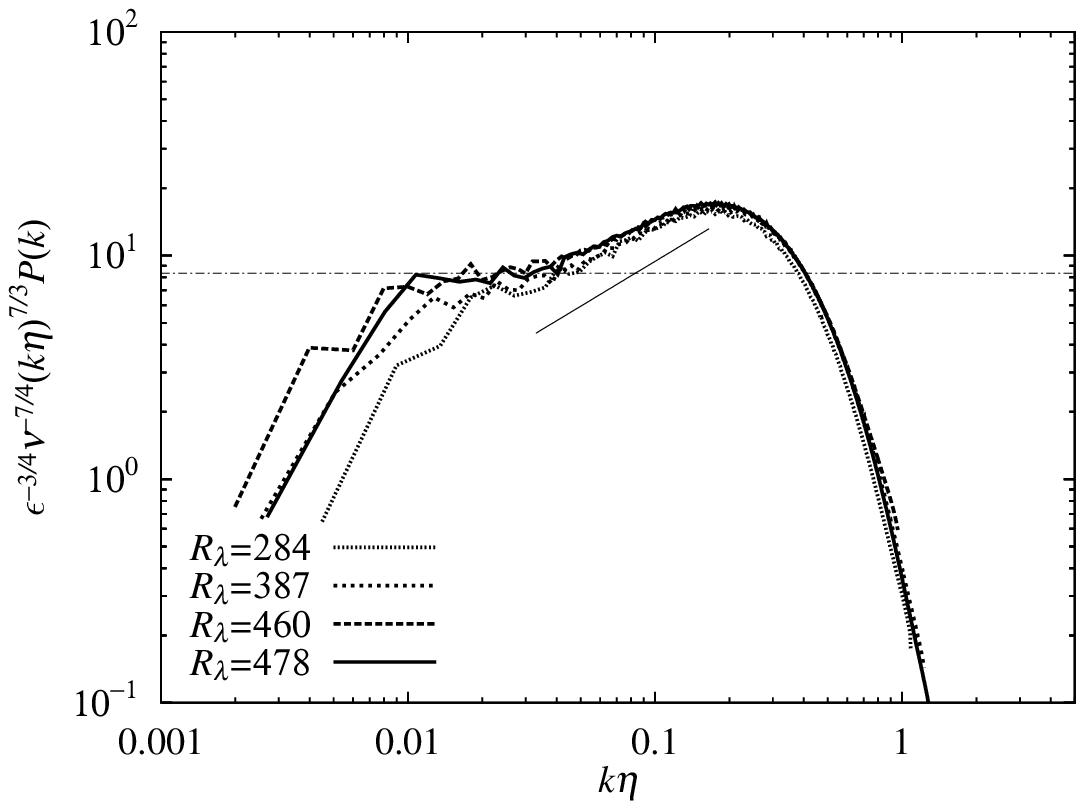}{Close up of
$\eb^{-3/4}\nu^{-7/4}(k\eta)^{7/3}P(k)$ for higher Reynolds numbers. A
short straight line shows the slope of $k^{2/3}$ and 
a horizontal line indicates $B_p=8.34.$}
\figline{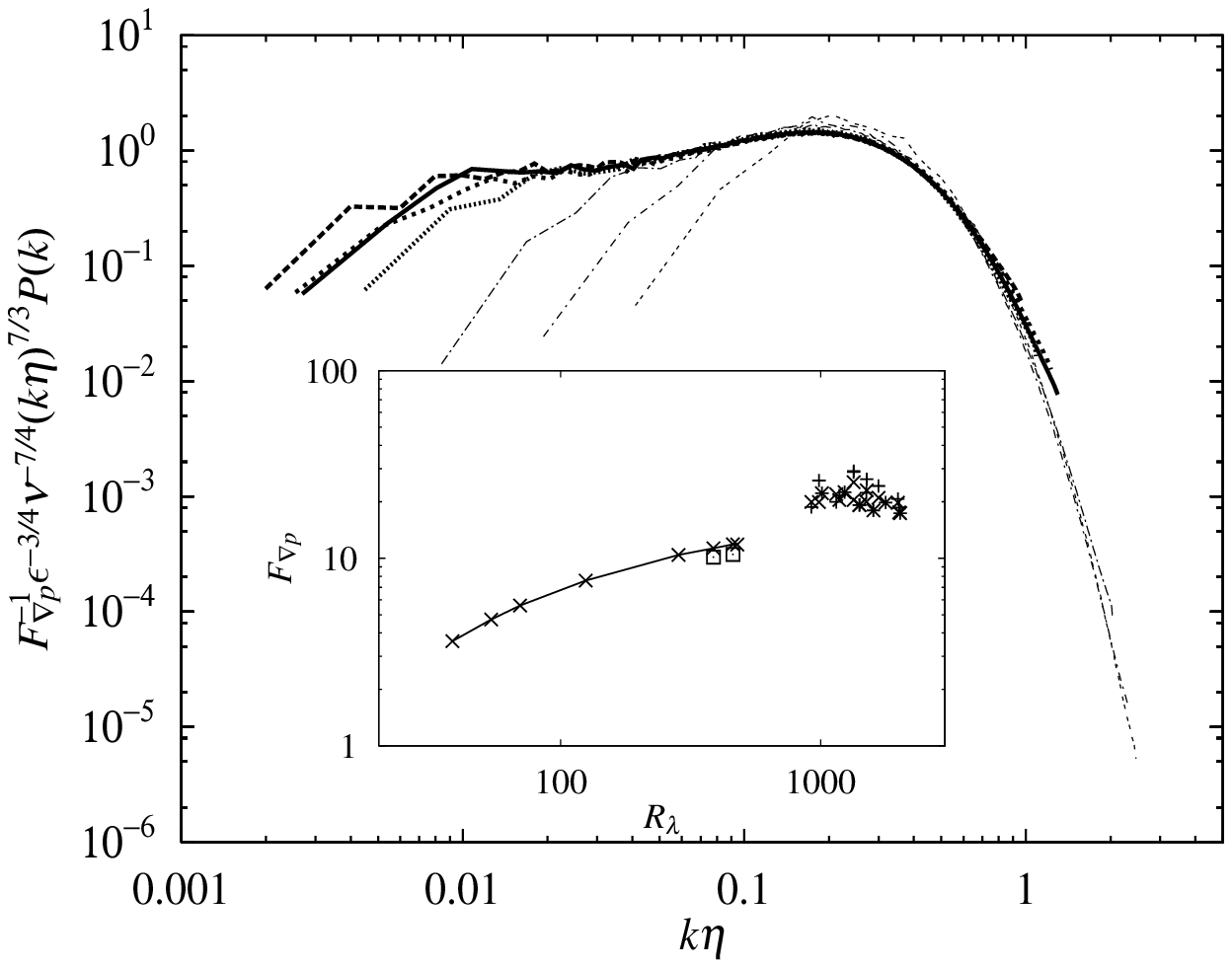}{Scaling of the pressure
spectra with the factor $F_{\nabla p}$, $F_{\nabla
p}^{-1}\eb^{-3/4}\nu^{-7/4}$ $(k\eta)^{7/3}P(k)$.  The lines are the same
as those in Fig. 2. The inset is the variation of $F_{\nabla p}$ 
against $\Rl$. Squares are $F_{\nabla p}$ 
computed by $\lim_{r\to 0} D_p(r)/r^2$, and other 
symbols are experimental data by Voth \etal\cite{Vothetal98}}

%%%%%% Structure functions %%%%%%
\figline{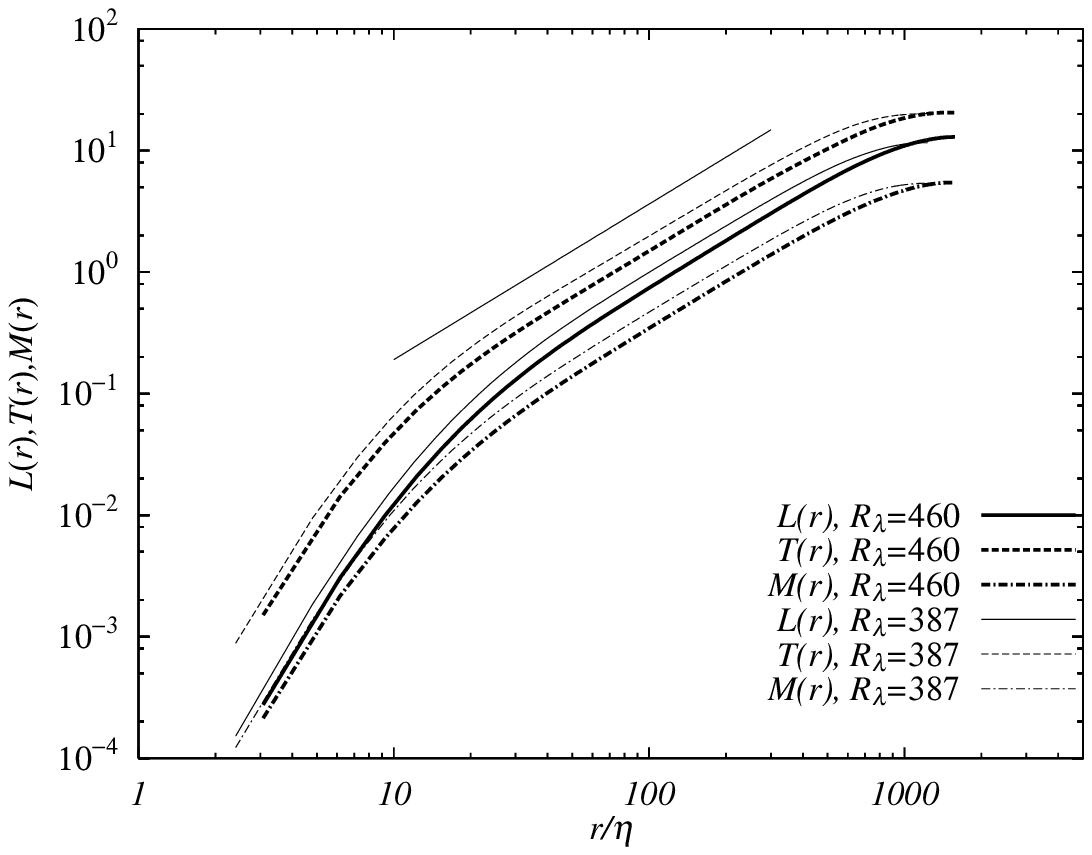}{Fourth order structure functions of velocity 
increments. $\Rl=387$ and $\Rl=460$ (thick lines). 
A straight line shows the slope of $1.28$.}
\figline{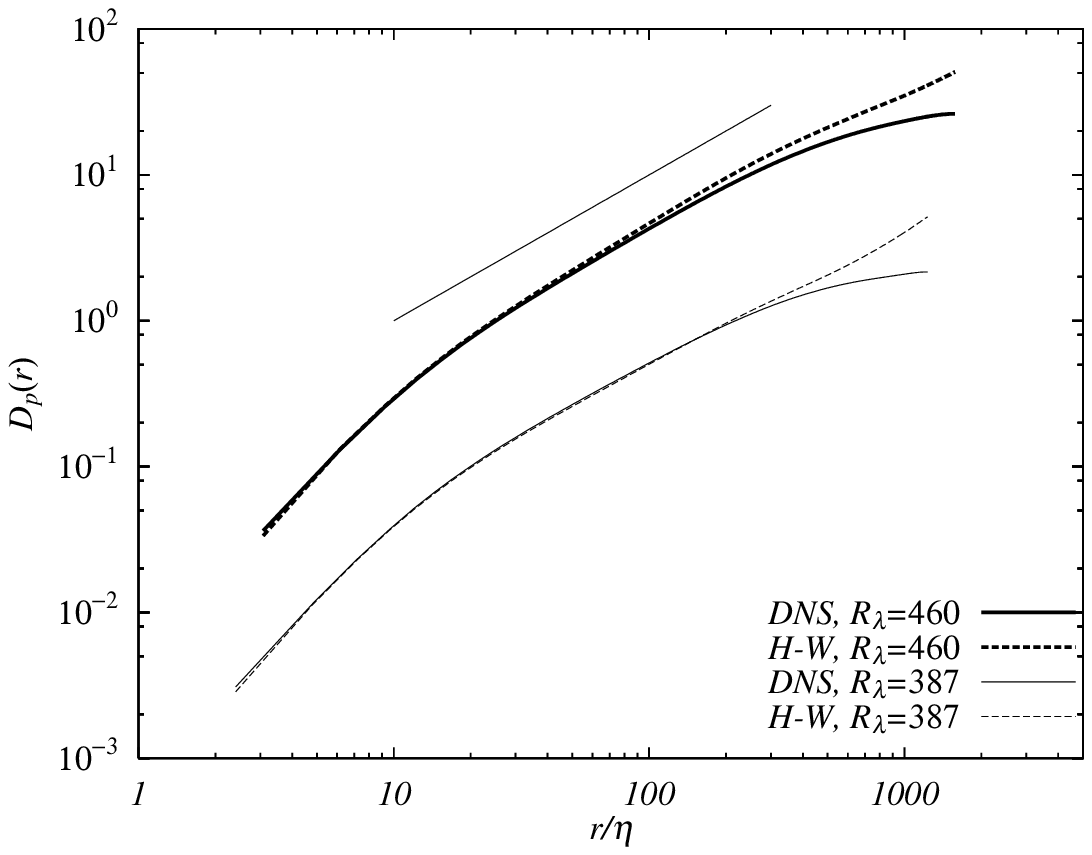}{Comparison of $D_p(r)$ with H \& W.\cite{Hw95}
 $\Rl=387$ and $\Rl=460$ (thick lines). A straight line shows the slope of
$r^1$. 
The curves for $\Rl=460$ are shifted upward by one unit for clarity. }

\vspace*{-0.5cm}
\end{multicols}

\end{document}